\documentclass[aip,jcp,amsmath,amssymb,floatfix,reprint]{revtex4-1}

\usepackage{graphicx}
\usepackage{dcolumn}
\usepackage{bm}

\usepackage{mathtools}
\usepackage{units}

\usepackage{color}

\usepackage{mathrsfs}
\usepackage{cases}
\usepackage{braket}
\usepackage{mathtools}

\usepackage[utf8]{inputenc}
\usepackage{hyperref}
\usepackage{breakurl}

\newcommand\mk[1]{\mkern-#1mu}

\newcommand\myp{'\mk2}
\newcommand\mypp{'\mk{2.5}'\mk2}

\newcommand\Lp{{\Lambda\mk2\myp}}
\newcommand\Lpp{{\Lambda\mk2\mypp}}

\newcommand\Jp{{J\mk1\myp}}
\newcommand\Jpp{{J\mk1\mypp}}

\newcommand\Vp{{v\myp\mk1}}
\newcommand\Vpp{{v\mypp}}

\newcommand\rar{\rightarrow}
\newcommand\srar{\mk3\rar\mk4}
\newcommand\LL{{\Lp \rar \Lpp}}
\newcommand\eff{\text{eff}}

\newcommand\dela[2]{\dfrac{\displaystyle #1}{\displaystyle #2}}

\newcommand\Gtun{\Gamma_{\mk3\Vp\Jp\Lp}^\text{tun}}
\newcommand\Grad{\Gamma_{\mk3\Vp\Jp\Lp \rar \Lpp}^\text{rad}}
\newcommand\Gtot{\Gamma_{\mk3\Vp\Jp\Lp\Lpp}^\text{tot}}

\newcommand\rturn{{r_{\mk3\curlywedge}}}

\begin{document}
%
\title[Radiative Association of CF$^+$]{Reaction Rate Constant for Radiative Association of CF$^+$}

\author{Jonatan Öström}
 \email[e-mail: ]{jonatan.ostrom@gmail.com}
 \affiliation{ Applied Physics, Division of Materials Science, Department of Engineering Science and Mathematics, Luleå University of Technology, 97187 Luleå, Sweden.}

\author{Dmitry S. Bezrukov}
\affiliation{Department of Chemistry, M. V. Lomonosov Moscow State University, Moscow 119991, Russia}

\author{Gunnar Nyman}
 \affiliation{Department of Chemistry and Molecular Biology, University of Gothenburg, 41296 Gothenburg, Sweden}%

\author{Magnus Gustafsson}%
 \email[e-mail: ]{magnus.gustafsson@ltu.se}
 \affiliation{ Applied Physics, Division of Materials Science, Department of Engineering Science and Mathematics, Luleå University of Technology, 97187 Luleå, Sweden.}

 \date{\today}

\begin{abstract}
\label{sec:abstract}
Reaction rate constants and cross sections are computed for the radiative association of carbon cations (C$^+$) and fluorine atoms (F) in their ground states. We consider reactions through the electronic transition $1^1\Pi \rightarrow X^1\Sigma^+$ and rovibrational transitions on the $X^1\Sigma^+$ and $a^3\Pi$ potentials. Semiclassical and classical methods are used for the direct contribution and Breit--Wigner theory for the resonance contribution. Quantum mechanical perturbation theory is used for comparison. A modified formulation of the classical method applicable to permanent dipoles of unequally charged reactants is implemented. The total rate constant is fitted to the Arrhenius--Kooij formula in five temperature intervals with a relative difference of $<3\:\%$. The fit parameters will be added to the online database KIDA. For a temperature of $10$ to $250\:$K, the rate constant is about $10^{-21}\:$cm$^3$s$^{-1}$, rising toward $10^{-16}\:$cm$^3$s$^{-1}$ for a temperature of $30{,}000\:$K. 
\end{abstract}
 
\maketitle
%
\section{introduction\label{sec:introduction}}
The fluoromethylidynium cation (CF$^+$) has been observed in the interstellar medium~\cite{neufeld-detect,neufeld-discover}. In a hydrogen abundant environment the major contribution to its production is the reaction $\rm HF + C^+ \rightarrow H + CF^+$, where HF is produced by $\rm H_2 + F \rightarrow HF + H$~\cite{neufeld-flourine,neufeld-halogens}. In this paper we investigate the possibility for production through radiative association of the reactants C$^+$ and F, which may be of importance in H$_2$-deficient environments.  

Radiative association may occur when at least one electronic state of a system of two reactants has a potential energy well below the dissociation energy. 
The system can reside in a bound state supported in this well if the collision and binding energies are expelled through the emission of a photon. 
The emission is due to the transition dipole moment or permanent electric dipole moment of the molecular complex during the collision. 
Magnetic and higher electric moments are not accounted for here, but do in general contribute. 
Radiative association of two fragments can be important in sparse interstellar gas where it can dominate over reactions due to many-body collisions\cite{Bates_51_MNRAS_111_303,Babb_Kirby_98} as the latter diminishes more rapidly with a decreasing number density of reactants. 

Modelling of the interstellar environment requires computation of the collision reaction rate\cite{Bates_51_MNRAS_111_303,Babb_Kirby_98}
\begin{equation}
r = k(T)[A][B]
\end{equation}
for all relevant species $A$ and $B$, which in turn requires the rate constant $k(T)$ for the species. 
In this paper we are concerned with finding the rate constant for the reaction $\text{C}^+ + \text{F} \rightarrow \text{CF}^+ + \hbar \omega$ through the three channels

\vspace{6pt}\noindent
$\text{C}^+(^2P) + \text{F}(^2P) \rar$
\begin{subnumcases}{\label{eq:reactions}}
\text{CF}^+(1^1\Pi) 		&$\rar\quad \text{CF}^+(X^1\Sigma^+) 	+ \hbar\omega $ \label{eq:1pi}\\
\text{CF}^+(X^1\Sigma^+) 	&$\rar\quad \text{CF}^+(X^1\Sigma^+) 	+ \hbar\omega $ \label{eq:1sig}\\
\text{CF}^+(a^3\Pi)		&$\rar\quad \text{CF}^+(a^3\Pi) 			+ \hbar\omega $.\label{eq:3pi}
\end{subnumcases}

Based on our electronic structure calculations (see Sec.~\ref{sec:abi}), we claim that the reactions (2) are the most important for the production of CF$^+$ through radiative association.  There are 12 electronic states correlating with ground state C$^+$ and F\cite{petsalakis2000}. Out of those only $X^1\Sigma^+$ and $a^3\Pi$ support bound states.  Of the remaining ten, $1^1\Pi$ allows for the closest approach.

The computational methods have been presented before, \textit{e.g.} in Refs.~\onlinecite{mgHF,mgCN,mgClassical,review} with the exception of changes to the classical theory to account for dipole moments that are non-zero at large separations. 
This paper is structured as follows. In Sec.~\ref{sec:theory} we outline the theory and numerical implementations of the computational methods. In Sec.~\ref{sec:abi} the \textit{ab initio} computations of potential energy curves and electric dipole moment curves are described. In Sec.~\ref{sec:results} the cross sections and rate constants for the three reaction channels are presented, as well as the fit of the total rate constant to the Arrhenius--Kooij formula. In Sec.~\ref{sec:conc} conclusions are drawn. 
\section{Methods and Theory\label{sec:theory}}
The reaction rate constant may be computed from the reaction cross section $\sigma_{\mk2\LL}(E)$ through
\begin{eqnarray}
k_\LL(T)
&=& \sqrt{\frac{8}{\mu \pi}} \left( \frac{1}{k_B T} \right)^{3/2} \nonumber\\
& &\times \int_0^\infty E\sigma_{\mk2\LL}(E) e^{-E/k_B T} dE 
\, ,
\label{eq:rate}
\end{eqnarray}
where $\mu = m_{\text{C}^+} m_\text{F} / (m_{\text{C}^+} + m_\text{F})$ is the reduced mass of the system, $k_B$ is Boltzmann's constant, $E$ is the collision energy, and $T$ is temperature. 
A single prime refers to the initial scattering state and a double prime to the final state. 
$\Lambda$ is the projection of the electronic orbital angular momentum onto the internuclear axis, and will in general denote different electronic states. 

A potential energy curve with a well may house bound vibrational states below the separation energy, and quasi-bound states above this energy. 
Also potentials that are monotonically approaching the separation energy from a single well, may support quasibound states due to the centrifugal barrier in the \textit{effective potential} when the molecule is rotationally excited. 
In quantum mechanical theory the reactants can tunnel in through the barrier and reside in a quasibound state, which is not classically accessible. 
The lifetime of the quasibound state for a collision energy $E$, which is related to the magnitude of the scattering wave function behind the barrier, strongly affects the energy dependent cross section, creating sharp peaks or \textit{resonances}; these features will be referred to as the \textit{resonance contribution} to the cross section or rate constant. 
Classical trajectories do not have this property and instead produce a smooth cross section, usually resembling a baseline of the spiky quantum mechanical dito; this will be referred to as the \textit{direct contribution}. 

The radiative association cross section for each reaction channel may be computed quantum mechanically for a grid of collision energies. 
It is proportional to the probability for the system to emit a photon due to its electric dipole moment and make a transition \textit{from} the scattering state of the given collision energy \textit{into} any bound state.
The cross section of this perturbation theory (PT) method is used here only for verification of the cross sections obtained using other methods. 
The reason for this is that unlike our other methods, PT produces a complete cross section including the direct \textit{and} the resonance contribution; but when there are narrow resonances it may not be reliable\cite{Bennett2003,radassHeH2plus}, and is therefore not used to produce the rate constant. 

Two methods are used that are based on classical trajectories.
The classical method (CL) rely on the Larmor power of the radiation from a time dependent dipole. 
The semiclassical method (SCL) is deduced from the semiclassical limit of the quantum mechanical optical potential method. 
Together they will be refered to as (S)CL. 
These methods produce only the direct contribution to which the resonance contribution
can be added by using Breit-Wigner (BW) theory.
The BW method requires the inverse lifetimes, or \textit{widths} of quasibound states. 
These are computed using the \textsc{Level} program \cite{leroy}. 
The BW cross section can be integrated analytically to produce a rate constant which may in turn be added to the classical dito.

\subsection{PT Method}
In PT the wave functions for the initial and final states must be obtained. Applying a partial wave expansion of the total wave function yields the ordinary time independent Schrödinger Equation 
\begin{equation}
\left( -\frac{\hbar^2}{2\mu} \frac{d^2}{dr^2} + V_\eff(r,J) \right) \Psi = E\Psi 
\, .
\label{eq:sch}
\end{equation}

For \underline{$E > 0$}, (with the energy in the dissociation limit $\equiv 0$) the scattering wave function $\Psi = F_{E\Jp}^\Lp$ is found for a number of equally spaced collision energies $E$, using the effective potential $V'_\eff(r,J')$ of the electronic state of approach $\Lp$ and the rotational quantum number $\Jp$. The wavefunction is energy normalized as in Ref.~\onlinecite{landau}. The integration of Eq.~\eqref{eq:sch} is in this case done with Numerov's method. The effective potential is constructed as
\begin{equation}
V_\eff(r,J) = V(r) + \frac{\hbar^2 J(J+1)}{2\mu r^2} 
\, ,
\end{equation}
where $V(r)$ is the \textit{ab initio} potential (see Sec.~\ref{sec:abi}) and the last term is the centrifugal energy.

For \underline{$E<0$}, Eq.~\eqref{eq:sch} is an eigenvalue problem on the target state effective potential $V''_\eff(r,J'')$. It is solved with the DVR method\cite{DVR-Lill-86,DVR-Light-85} for the bound wave functions $\Psi = \Psi_{\Vpp\Jpp}^{\Lpp}$, which are normalized to unity. $v''$ is the vibrational quantum number. 

The Einstein A-coefficient for spontaneous emission from the scattering state $a$ to the bound state $b$ is derived from the perturbation Hamiltonian that couples the electromagnetic field of the emitted photon to the molecular dipole $\mathbf{D}$ under the dipole approximation; it can be written as~\cite{review}
\begin{align}
A_{ab} &= \frac{k_e}{\hbar} \frac{32\pi^3}{3} \frac{|\mathbf{D}_{ab}|^2}{\lambda_{ab}^3}
\, , \\
\intertext{and can be turned into a cross section}
\sigma_{ab}(E) 
& = \pi^2\hbar^3 \frac{P_\Lp}{\mu E}  A_{ab} \nonumber\\
& = k_e \hbar^2 \frac{32\pi^5}{3} \frac{P_\Lp}{\mu E} \frac{|\mathbf{D}_{ab}|^2}{ \lambda^3_{ab}}
\, ,
\end{align}
where $k_e = (4\pi\epsilon_0)^{-1}$ is Coulomb's constant, $P_\Lp$ is the probability of approach in state $\Lambda'$, and
\begin{equation}
\lvert\mathbf{D}_{ab}\rvert^2 = \mathscr{S}_{\Lp\Jp,\Lpp\Jpp} \lvert\braket{F_{E\Jp}^{\Lp}(r)|D_{\Lp\Lpp}(r)|\Psi_{\Vpp\Jpp}^{\Lpp}(r)}\rvert^2
\, .
\end{equation} 
The Hönl--London factors\cite{honlLondon} $\mathscr{S}_{\Lp\Jp,\Lpp\Jpp}$ are drawn from Ref.~\onlinecite{honl} and are listed with $P_\Lp$ in Table~\ref{tab:honl} for each transition. $D_{\Lp\Lpp}(r)$ is the \textit{ab initio} electric dipole moment (see Sec.~\ref{sec:abi}). 

With $\lambda_{ab} = \lambda_{E\Lpp\Vpp\Jpp}$, summation over all lower vibrational and allowed rotational levels gives the total cross section
\begin{eqnarray}
\sigma_\LL (E) 
&=& k_e \hbar^2 
\frac{32\pi^5}{3}
\frac{P_\Lp}{\mu E}   \sum_{\Jp;\Vpp,\Jpp} 
\frac{\mathscr{S}_{\Lp\Jp,\Lpp\Jpp}}{\lambda_{E\Lpp\Vpp\Jpp}^3} \nonumber \\ 
&& \times\, \lvert\braket{F_{E\Jp}^\Lp(r)|D_{\Lp\Lpp}(r)|\Psi_{\Vpp\Jpp}^{\Lpp}(r)}\rvert^2 
\, .
\end{eqnarray}
\begin{table}
\centering
\renewcommand\arraystretch{1.2}
\caption{Hönl--London factors, $\mathscr{S}_{\Lp \Jp , \Lpp \Jpp}$, and statistical weights,
  $P_{\Lp}$, for CF$^+$.
  The Hönl--London factors are parity averaged for the case
  $1^1\Pi \rightarrow X^1\Sigma^+$.\label{tab:honl}}
\begin{tabular}{@{}lllll@{}}
 & $\mathscr{S}_{\Lp \Jp , \Lpp (\Jp-1)}$ & $\mathscr{S}_{\Lp \Jp , \Lpp \Jp}$ & $\mathscr{S}_{\Lp \Jp , \Lpp (\Jp+1)}$ & $P_\Lp$ \\
\hline\hline
$1^1\Pi \srar  X^1\Sigma^+$ & $(J'+1)/2$ & $(2J'+1)/2$ & $J'/2$ & $2/36$ \\
$X^1\Sigma^+$ & $J'$ & $0$ & $J'+1$ & $1/36$ \\
$a^3\Pi$ & $\frac{(J'+1)(J'-1)}{J'}$ & $\frac{2J'+1}{J'(J'+1)} $ & $\frac{J'(J'+2)}{(J'+1)}$ & $6/36$ \\
\hline
\end{tabular}
\end{table}

\subsection{BW Method}
According to Heisenberg's uncertainty principle $\Delta E \Delta t \geq \hbar/2$, the finite lifetime $\tau = \Delta t$ of a quasibound state determined by $v'J'\Lambda'$ at energy level $E_{\Vp\Jp\Lp}$, corresponds to the total \textit{width} $\Gtot \equiv 2\Delta E = \hbar/\tau$. 
%
%
This state can dissociate by tunneling back through the barrier or a photon can be emitted resulting in a transition into any lower-lying level. We set
\begin{equation}
\Gtot = \Gtun + \Grad
\, ,
\end{equation}
where $\Gtun$ is the tunneling width and $\Grad$ is the radiative width corresponding to a transition into a bound or lower-lying quasibound state, thereby neglecting other processes (such as predissociation or radiative transitions into lower-lying free states). 

The BW cross section is \cite{review}
\begin{equation}
\sigma_\LL (E)  =
\frac{\pi\hbar^2}{2\mu E} P_\Lp
\sum_{\Vp\Jp} 
\dela{ (2J'+1) \, \Gtun \, \Grad }
{(E-E_{\Vp\Jp\Lp})^2 + (\Gtot/2 )^2 }
\, ,
\end{equation}
and may be integrated analytically in Eq.~\eqref{eq:rate} by assuming for each resonance that $\Gtot \ll E_{\Vp\Jp\Lp}$ so that $e^{-E_{\Vp\Jp\Lp}/k_B T}$ may replace $e^{-E/k_B T}$. 
The resulting expression can be written
\begin{eqnarray}
k_\LL (T) 
& = & \hbar^2 \left(\frac{2\pi}{\mu k_BT}\right)^{3/2} P_\Lp \nonumber\\
& &\times \sum_{\Vp\Jp}
\dela{ (2J'+1)\mk{-2}\, e^{ \mk3 \textstyle \frac{ \scriptstyle -E_{\Vp\Jp\Lp} }{ \scriptstyle k_B T} } }
{1/{\Gtun} + 1/{\Grad}}
\, .
\end{eqnarray}

The BW method requires the knowledge of $\Grad$, $\Gtun$ and $E_{\Vp\Jp\Lp}$ for all quasibound states. 
These were found with the computer program \textsc{Level} 8.0 \cite{leroy}. 
The program did not perform well for the double minima in the effective potentials for reaction channels \eqref{eq:1pi} and \eqref{eq:3pi}. In these cases the radial distance was divided into two overlapping intervals, each containing one of the minima. 
The cross section produced in this way closely resembles that from PT, which supports the taken approach. 

\subsection{SCL Method}
The SCL method~\cite{Bates_51_MNRAS_111_303,mgSiN} is derived as the semiclassical limit of the cross section of the distorted wave optical potential method~\cite{zyg-dal-semiclassical-88,mies69} by assuming small phase shifts and applying the WKB approximation \cite{review}. 
The SCL method is applicable only to radiative association involving an electronic transition (reaction \eqref{eq:1pi} in this case). 
The cross section is
\begin{equation}
\sigma_\LL (E) 
 = 4\pi 
\sqrt{\frac{\mu }{2}} P_\Lp 
\int\limits_{0}^\infty b 
\int\limits_\rturn^\infty 
\frac{A_\LL^{Eb}(r)}
{ \sqrt{E - V'_\eff(r,b,E)} } drdb
\, ,
\label{eq:scl}
\end{equation}
where $b$ is the impact parameter, \textit{i.e.} the asymptotic offset from a head on collision, $\rturn$ is the classical turning point and 
\begin{align}
A_\LL^{Eb}(r) &=
\left\lbrace
\begin{array}{ll}
	A_\LL(r) &
		\begin{array}{ll}
		\texttt{if } & E < V'(r)-V''(r) \\
		\texttt{and } & V''_\eff(r,b,E)<0\, ,
		\end{array} \\
	0 & \begin{array}{l} \texttt{else}\, , \end{array}
\end{array}
\right. 
\\
A_\LL(r) &= 
\frac{k_e}{\hbar} 
\frac{32\pi^3}{3} 
\mk{-6}
\underbrace{\mk6
\left( 
 \frac{2-\delta_{0,\Lp + \Lpp}}
	{2-\delta_{0,\Lp}}
\right)\mk6
}_{\mk{30}\phantom{===}=1 \text{ for reactions \eqref{eq:reactions}}\mk{30}}
\mk{-6}
\frac{D_{\Lp\Lpp}^2(r)}
	{\lambda_{\Lp\Lpp}^3(r)}
\, .
\end{align}
The effective potentials and the optimal wavelengths are constructed as 
\begin{align}
V_\eff(r,b,E) &= V(r) + Eb^2/r^2 
\, ,\\
\lambda_{\Lp\Lpp}(r) &= \frac{2\pi\hbar c}{V'(r) - V''(r)}
\, .
\end{align}

This cross section is smooth and can be reliably integrated in Eq.~\eqref{eq:rate} and added to the BW result. 
Romberg integration is used for the $r$ integral in Eq.~\eqref{eq:scl}, and the trapezoidal rule  for $b$. Simpson's \nicefrac{1}{3} rule is used for the $E$ integral in Eq.~\eqref{eq:rate}, and $\rturn$ is found using bisection.

\subsection{CL Method}
The CL theory is based on classical trajectories and the Larmor power~\cite{jackson} radiated by a time dependent dipole~\cite{levine}. 
The method applies only to non-electronic transitions, \textit{i.e.} reactions \eqref{eq:1sig} and \eqref{eq:3pi}  in our case.
A generalization of the resonance free cross section derived in Ref.~\onlinecite{mgClassical} is
\begin{equation}
\sigma_\Lambda(E) 
= \frac{k_e}{\hbar} \frac{4 P_\Lambda}{3c^3} 
 \int\limits_0^\infty  b
\int\limits_{E/\hbar}^{ \omega_{\text{max}} }  \frac{1}{\omega} 
\Bigg\lvert \int\limits_{-\infty}^{\infty}  \ddot{\mathbf{D}}(b,E,t) e^{i\omega t} dt \Bigg\rvert^2 
d\omega db 
\, ,
\label{eq:classical}
\end{equation}
where $\hbar \omega_{\text{max}} = E - \min(V_\eff)$ is the maximum photon energy that is possible between the collision energy and the absolute minimum of the effective potential. 
Applying the Fourier transform derivative property 
\begin{equation}
\lvert\mathscr{F}(\ddot{D})\rvert^2 = \omega^4\lvert\mathscr{F}(D)\rvert^2 
\, ,
\label{eq:fourier}
\end{equation}
for asymptotically vanishing functions $D(t\rar\pm\infty)=0$, would yield the expression in Ref.~\onlinecite{mgClassical}. Since the permanent dipoles (see Fig.~\ref{fig:pot_dip}) asymptotically approaches the dipole moment given by the position of the charged reactant C$^+$ relative to the systems center of mass, \textit{i.e.} 
\begin{equation}
D_{r \rar \infty} = -q_er \frac{ \raisebox{-2pt}{$m_\text{F}$} }{ \raisebox{5pt}{$m_{\text{C}^+} + m_\text{F}$} } 
\, ,
\label{eq:asymp_dip}
\end{equation}
Eq.~\eqref{eq:fourier} does not hold. Instead the squared expression in Eq.~\eqref{eq:classical} is evaluated as follows. (The arguments $(b, E, t)$ of variables $D$, $\mathbf{D}$, $D_x$, $D_y$, $r$ and $\theta$ are omitted for conciseness.) 
The time dependent dipole
\begin{align}
\mathbf{D} 
&= \bigg[\begin{array}{l} D_x \\D_y \end{array} \bigg]= 
\biggl[
 \begin{array}{l}
 D \cos \theta\\
 D \sin \theta
 \end{array} \biggr]
\\
\intertext{is obtained by integrating the equations of motion }
\dot{r} 
&= \sqrt{\frac{2}{\mu} \left( E - V_\eff(r,b,E) \right)} \label{eq:R}\\
\dot{\theta} 
&= \frac{b}{r^2} \sqrt{\frac{2E}{\mu}} 
\, , \label{eq:theta}
\end{align}
where the angle $\theta$, being defined as
the orientation of the molecule, gives the dipole's direction.
The second time derivative of the dipole is
\begin{equation}
\ddot{\mathbf{D}} = 
\bigg[\begin{array}{l} \ddot{D}_x \\ \ddot{D}_y \end{array}\bigg] =
\bigg[
 \begin{array}{l}(\ddot{D} - D \dot\theta^2) \cos\theta
- (2 \dot{D} \dot\theta + D \ddot\theta)\sin\theta \\
(\ddot{D} - D \dot\theta^2) \sin\theta
+ (2 \dot{D} \dot\theta + D \ddot\theta)\cos\theta
 \end{array} 
 \bigg]
 \label{eq:Ddd}
\end{equation}
where Eq.~\eqref{eq:theta} yields
\begin{equation}
\ddot\theta = -\frac{\dot{r}}{r^3}2b\sqrt{\frac{2E}{\mu}}
\, ,
\end{equation}
and the first time derivatives $\dot{r}$ and $\dot\theta$ are readily available in the numerical implementation.
In the coordinate system of Ref.~\onlinecite{McQuarrie68} $r(t=0) = \rturn$ and $\theta(t=0)=0$.
Then $\ddot{D}_x$ is symmetric in time and
$\ddot{D}_y$ is anti-symmetric.
The squared Fourier transform of the
dipole can thus	 be computed as
%
%
\begin{eqnarray} \label{eq:cos_sin}
\biggl\lvert\int_{-\infty}^\infty \ddot{\mathbf{D}} e^{i \omega t} dt \biggr\rvert^2 
& = & \biggl(2 \int_0^\infty \cos(\omega t)  \ddot{D}_x dt \biggr)^2 \nonumber\\ 
& & \mkern3mu \mathrel{+}
\bigg(2 \int_0^\infty \sin(\omega t) \ddot{D}_y dt \bigg)^2 
\, .
\end{eqnarray}

When computing Eq.~\eqref{eq:cos_sin} the derivatives $\dot{D}$ and $\ddot{D}$ in Eq.~\eqref{eq:Ddd} are evaluated with finite difference with the same time step as the fourth order Runge-Kutta integration of the trajectory. The Fourier transform is carried out with sine and cosine FFTs from Ref.~\onlinecite{numRes}. The $\omega$ integral in Eq.~\eqref{eq:classical} is computed with Simpson's \nicefrac{1}{3} rule. Integrating over $b$ and $E$ and finding $\rturn$ is done as in the SCL case. 
\section{\label{sec:abi} Molecular Potentials and Dipole moments}
Data points for the potential energy curves (PECs) and  permanent and transition electric dipole moment curves (DMCs) were determined with \textit{ab initio} electronic structure calculations. The data points were inter- and extrapolated to give smooth functions for the required range in internuclear distance. 

\subsection{Ab Initio Electronic Structure Calculations}  
All calculations were performed for internuclear distances from 1.5 to 7.0 $a_0$ in steps of 0.1 $a_0$. 
The molecular orbitals were constructed using the CASSCF method with an active space consisting of 10 electrons in 8 orbitals, which at the dissociation limit correspond to the 2s and 2p orbitals of the separate atoms. 
The averaging was done over all 36 components corresponding to 12 electronic states ($X^1\Sigma^{+}$, $1^1\Sigma^{+}$, $1^1\Sigma^{-}$, $1^1\Pi$, $2^1\Pi$, $1^1\Delta$,$1^3\Sigma^{+}$, $2^3\Sigma^{+}$, $1^3\Sigma^{-}$, $a^3\Pi$, $2^3\Pi$, $1^3\Delta$)  correlating with the lowest dissociation limit of the system: C$^{+}({^2}P)+\text{F}({^2}P)$. 
Then the PECs and the corresponding DMCs were calculated with the internally contracted MRCI method with Davidson correction using the CASSCF molecular orbitals as a reference. 
The calculations were carried out with aug-cc-pV5Z and aug-cc-pV6Z Dunning-type basis sets using the standard contraction scheme. 
Furthermore, a calculation was performed with the aug-cc-pV5Z-DK basis set. 
In this case, the scalar relativistic correction was accounted for by the second order Douglas--Kroll--Hess (DKH) Hamiltonian. 
All calculations were carried out in the $C_{2v}$ symmetry group. The MOLPRO 2010.1 package was used.

\begin{figure}[hbt]
\centering
\input{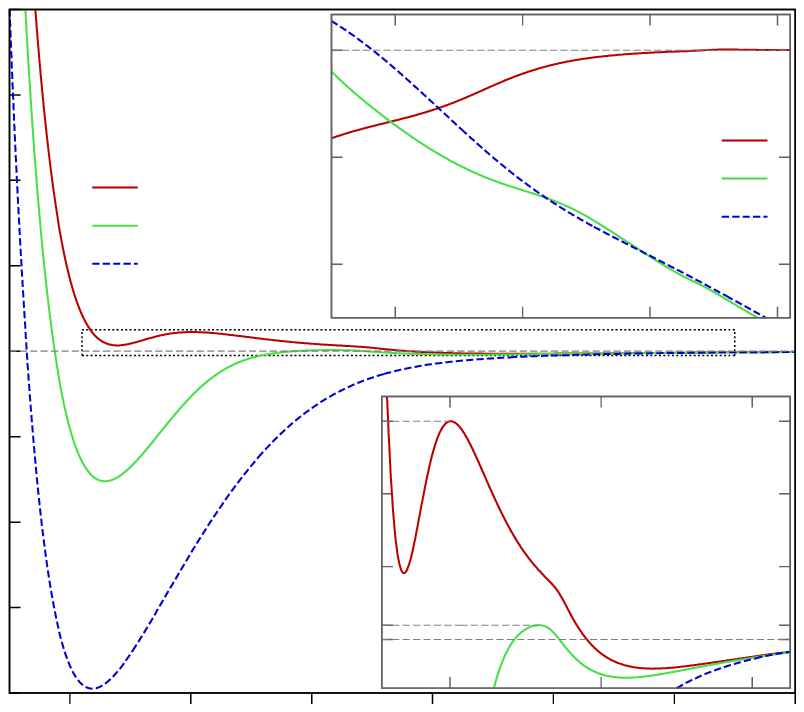}
\caption{\label{fig:pot_dip} The \textit{ab initio} potential energy curves of the three electronic states in reaction~\eqref{eq:reactions}.
\textsc{lower inset}: The content of the rectangle in the main plot, where potential barrier heights are indicated.
\textsc{upper inset}: The transient dipole moment of the electronic transition  in reaction channel~\eqref{eq:1pi} and the permanent dipole moments of~\eqref{eq:1sig} and~\eqref{eq:3pi}.}
\end{figure}

Estimating the PECs in the complete basis set (CBS) limit from the aug-cc-pV5Z and aug-cc-pV6Z calculations and adjusting for the scalar relativistic correction, was carried out using the extrapolation formula in the same manner as in Ref.~\onlinecite{Shi2012536}. 
The scalar relativistic correction was estimated by the difference between the aug-cc-pV5Z and aug-cc-pV5Z-DK calculations (it should be noted that in our calculation this correction does not exceed $100\:$cm$^{-1}$ in the interaction region). 
Identical DMC results were obtained in all three basis sets, and the aug-cc-pV6Z result was used. The calculated PECs and DMCs are shown in Fig.~\ref{fig:pot_dip}. 

\subsection{Inter- and Extrapolation of Ab Initio Data}
The extrapolation toward zero and infinity of the \textit{ab initio} data was done using the two first and two last data points respectively (as seen from the left in Fig.~\ref{fig:pot_dip}). 
PECs were extrapolated toward zero by the function $ V_{\text{min}} + Ae^{-\alpha r} $, where $V_{\text{min}}$ is the lowest data value for the potential. 
Extrapolation toward infinity of the $1^1\Pi$~PEC used the function
\begin{equation}
V_{r \rar \infty}^{1^1\Pi} (r) = -3.49/2r^4 - c_6/r^6 + \Delta E^{1^1\Pi}\:[\text{a.u.}]
\, ,
\end{equation}
which assumes a long range polarizability constant\cite{Zatsarinny09} $\alpha=3.49\:a_0^3$ for F, a dispersion term and an arbitrary energy offset in order to make the energy in the dissociation limit zero. 
To keep the relative difference between the potentials, the $a^3\Pi$ and $X^1\Sigma$ potentials reused the energy offset $\Delta E^{1^1\Pi}$ and a term $-c_8/r^8$ was added.

As no assumptions could be made about the behaviour of the DMCs between $0$ and $1.5\:a_0$, the extrapolation toward zero was the straight line connecting the first two data points. 
This should be safe (cf. Fig.~\ref{fig:crossCLQM}) as the classical turning points for potentials $X^1\Sigma^+$ and $a^3\Pi$ at $E=1\:e$V are located at $\rturn =1.63$ and $1.78\:a_0$ respectively, and at $\rturn = 1.66\:a_0$ for $1^1\Pi$ at $E=10\:e$V. Those energies are roughly the maximum relevant collision energies for each molecular state (see Sec.~\ref{sec:results}).
Toward infinity the $1^1\Pi \rar X^1\Sigma$ transition DMC was extrapolated with the function $Ae^{-\alpha r}$ and the permanent DMCs with $ -r m_\text{F}/(m_{\text{C}^+} + m_\text{F}) + Ae^{-\alpha r}$\:[a.u.] where the first term comes from Eq.~\eqref{eq:asymp_dip}.

A cubic spline with the endpoint derivatives acquired from the extrapolation was used for interpolation.
\section{Results\label{sec:results}}
\begin{figure}[!htb]
\centering
\input{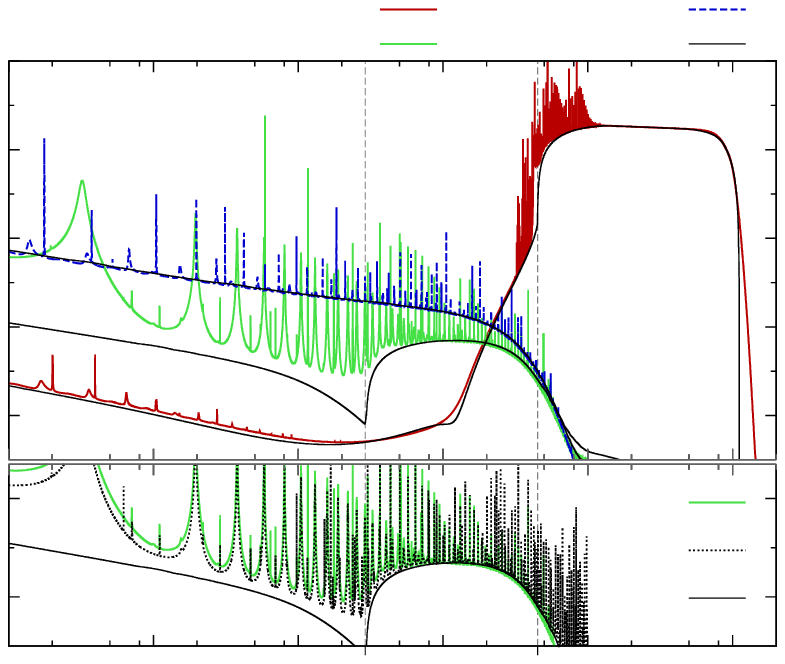}
\caption{\label{fig:crossCLQM} \textsc{upper panel:} Cross sections for reactions~\eqref{eq:reactions} from PT in color and from (S)CL in black. \textsc{lower panel:} Comparison of cross sections from CL+BW and PT approaches for transitions on $a^3\Pi$.}
\end{figure}

Here we present the numerical results for the cross section and the rate constant for the formation of CF$^+$ through the reactions \eqref{eq:reactions}. 
The resulting cross sections from (S)CL and PT are shown in the upper panel of Fig.~\ref{fig:crossCLQM}. 
Here it is apparent that the (S)CL cross sections resemble baselines of those obtained  with PT. 
The smaller the colliding species the more quantum mechanical they are in nature, but, apart from the resonance structure, C$^+$ and F seem appropriately large for roughly a 5\% accuracy with the (S)CL methods. 

We note that the dip in the $a^3\Pi$ cross section at $0.029$\:$e$V and the steep onset of the $1^1\Pi\rightarrow X^1\Sigma$ cross section at $0.45\:e$V correspond to the barrier heights of the corresponding potentials in Fig.~\ref{fig:pot_dip}. 
The $X^1\Sigma$ potential lacks a barrier and therefore has a smooth, monotonically decreasing baseline. 
The PT and (S)CL+BW cross sections for transitions on $a^3\Pi$ are shown in the lower panel of Fig.~\ref{fig:crossCLQM}. 
This is the reaction channel where these two methods yield the greatest relative difference. 
Still, the similarity between the cross sections produced by these two methods supports the general approach taken in the present study, \textit{i.e.} computing the rate constant as the sum (S)CL+BW. 

\begin{figure}[!htb]
\centering
\input{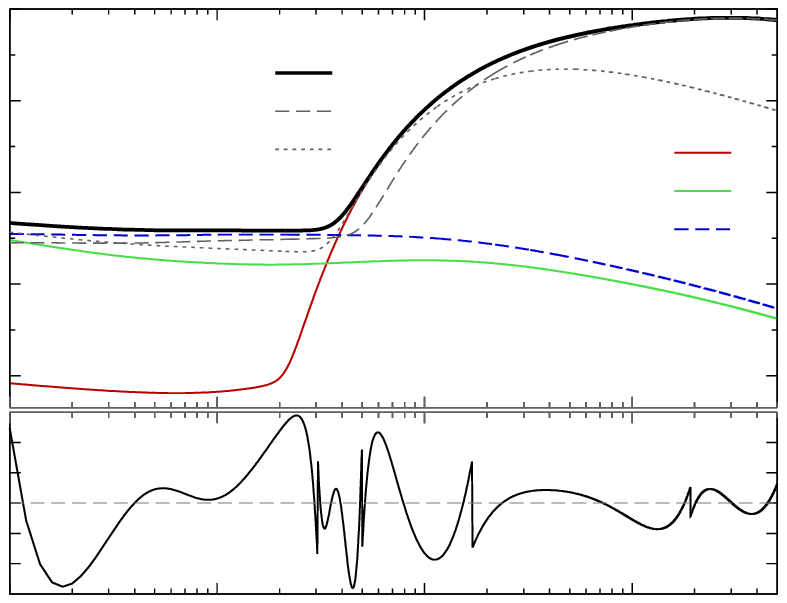}
\caption{\label{fig:rate} \textsc{upper panel:} Rate constants for reactions~\eqref{eq:reactions} from (S)CL+BW. The total sum is shown in black. For comparison, strictly (S)CL and strictly BW contributions to the total rate constant are shown as dashed grey lines. \textsc{lower panel:} The relative difference of the Arrhenius--Kooij fits and the computed rate constant.}
\end{figure}

The reaction rate constant is shown in the upper panel of Fig.~\ref{fig:rate}, where strictly (S)CL and strictly BW rate constants are also included for the comparison of the direct and the resonant contribution. 
The resonance mediated rate constant dominates over the direct for ${T<}\:20$  and $400\:{<T<}\:1100\:$K. 
This appears to be due to the low energy resonances housed behind the $0.029\:e$V $a^3\Pi$ barrier, and the ${\sim}0.45\:e$V resonances housed in the $1^1\Pi$ upper well, respectively. 
At $T\approx560\:$K the BW result is nearly six times  that of the (S)CL. 
The strictly rovibrational transitions of reaction channels~\eqref{eq:1sig} and~\eqref{eq:3pi} dominate at low temperatures up to $T\approx400\:$K with a combined rate constant $k\approx10^{-21}\:$cm$^3$s$^{-1}$. 
For increasing temperatures the electronic transition of channel~\eqref{eq:1pi} rapidly dominates with $k$ peaking below $10^{-16}\:$cm$^3$s$^{-1}$ at $T\approx30{,}000\:$K. 
This is qualitatively similar to other systems like CO with a barrier on the upper state potential that suppresses the low energy cross section; see \textit{e.g.} Ref.~\onlinecite{mgCO}. 

The rate constant was fitted to the Arrhenius--Kooij formula 
\begin{equation}
k(T) = A \left( T/300 \right)^B e^{-C/T}
\label{eq:kooij}
\end{equation}
in five intervals to adhere to the KIDA \cite{kida} database. The fit is very close to the total rate constant in the upper panel of Fig.~\ref{fig:rate}. The difference in percent can be seen in the lower panel. The fit parameters are listed in Table~\ref{tab:kooij}. 
\begin{table}[!htb]
\centering
\caption{\label{tab:kooij} Arrhenius--Kooij fit parameters for Eq.~\eqref{eq:kooij}}
\begin{tabular}{@{}lclclcl@{}} 
\hline\hline 
$T$ range $[\mkern 2mu$K$\mkern 2mu]$ && $A\:[\mkern 2mu 1\text{e}{-17}\mkern 2mu]$ && $B$ && $C$ \\ 
\hline 
$10 \to 305 		$	&& $0.000149002 	$	&& $0.063154	$	&& $-6.06328	$\\
$305 \to 500		$	&& $1.47468\text{e}{-13} $	&& $20.4233	$	&& $-6237.85	$\\
$500 \to 1 700		$	&& $0.571101 		$	&& $0.90067 	$	&& $3286.36		$\\
$1700 \to 19100 	$	&& $3.70502 		$	&& $0.174208 	$	&& $4372.76		$\\
$19100 \to 50000 	$	&& $683.555 		$	&& $-0.839364 	$	&& $23804.8 	$\\
\hline 
\end{tabular}
\end{table}

\section{conclusion\label{sec:conc}}
The production of CF$^+$ through radiative association has been studied.  Cross sections and rate constants have been computed with a combination of classical and quantum mechanical methods. 
The previously published\cite{mgClassical} classical (CL) theory has been modified to account for permanent dipoles of unequally charged reactants. The formula appears to work as well as the corresponding formula for radiative association of equally charged diatoms. 

The rate constant and cross section for the radiative association of CF$^+$ was computed. 
In Ref.~\onlinecite[(Table 4)]{neufeld-flourine} the reactions
\begin{subequations}
\begin{align}
 &\rm F + H_2 \rightarrow \rm HF + H,&	k &> 1.0\text{e}{-10}\:\text{cm}^3\text{s}^{-1}, \label{eq:FH2-HF}\\
 &\rm C^+ + HF \rightarrow \rm CF^+ + H,&	k &> 7.2\text{e}{-9}\:\text{cm}^3\text{s}^{-1}, \label{eq:CpHF-CFp}
\end{align}
\end{subequations}
are listed with their corresponding rate constants. These values should be compared with the total rate for radiative association in Fig.~\ref{fig:rate}. Assuming a vast abundance of H$_2$, the reaction~\eqref{eq:FH2-HF} should out-compete reactions~\eqref{eq:reactions} for the reactant F, and \eqref{eq:FH2-HF} should be the major contributor to the relative abundance of HF. Reaction~\eqref{eq:CpHF-CFp} should in turn be the major source of interstellar CF$^+$. The radiative association of CF$^+$ may be of importance in environments where H$_2$ is less abundant, for instance in metal rich ejecta of supernovae, similar to what has been concluded for the production of CO\cite{Petu1989,Gearhart1999ApJ}.

\begin{acknowledgments}
We would like to thank Nikolay V. Golubev for valuable discussions in the initial phase of this project, and for preliminary computational results. 
We acknowledge the support from the Swedish Research Council and the COST Actions CM1401 "Our Astro-Chemical History" and CM1405 "Molecules In Motion" (MOLIM). 
Part of the work related to the \textit{ab initio} calculations has been undertaken using the supercomputer facilities of the MSU research computer center and was supported by RFBR (project 14-03-00422).
We also acknowledge the free software used in this project, including GNU/Linux, GNU Fortran compiler, GNU Octave, Gnuplot, and the \TeX\ ecosystem. 
\end{acknowledgments}
\bibliography{citation}
\end{document}